\documentclass[twocolumn,showpacs,prl,superscriptaddress]{revtex4}
\usepackage{graphicx}
\usepackage{longtable}

\begin{document}

\title{Pressure- and temperature-induced structural phase transitions
of CaFe$_2$As$_2$ and BaFe$_2$As$_2$ studied in the Hund's rule
correlation picture}

\author{Wei Ji}
\affiliation{Department of Physics, Renmin University of China,
Beijing 100872, China}

\author{Xun-Wang Yan}
\affiliation{Institute of Theoretical Physics, Chinese Academy of
Sciences, Beijing 100190, China} \affiliation{Department of Physics,
Renmin University of China, Beijing 100872, China}

\author{Zhong-Yi Lu}
\affiliation{Department of Physics, Renmin University of China,
Beijing 100872, China}

\begin{abstract}

With the proposed Hund's rule correlation picture, i.e. the
fluctuating Fe local moments with the As-bridged antiferromagnetic
superexchange interactions, the exceptional collapsed tetragonal
phase and related phase transitions observed in CaFe$_2$As$_2$ are
well understood. With the same framework, a pressure-temperature
phase diagram is predicted for BaFe$_2$As$_2$ as well, in which a
paramagnetic tetragonal and a collinear antiferromagnetic
orthorhombic structures to nonmagnetic tetragonal structure
transitions take place around 4-8 GPa and 10-15 GPa respectively,
and a nonmagnetic tetragonal to a nonmagnetic collapsed tetragonal
structure transition takes place over 26 GPa. Our study helps better
understand the complex correlation among crystal structure,
magnetism, and electronic structure in pnictides, a precondition to
understand the superconductivity in pnictides.

\end{abstract}

\received[Dated: ]{\today }
\startpage{1}
\endpage{}
\pacs{74.70.Xa, 74.20.Pq, 74.25.Dw, 74.25.Bt}

\maketitle

Since the discovery of superconductivity in LaFeAsO by partial
substitution of O with F atoms below 26K\cite{kamihara}, intense
studies have been devoted to physical properties of iron-based
pnictides both experimentally and theoretically. There are, so far,
four types of iron-based compounds reported, showing
superconductivity after doping or under high pressures, i.e.
1111-type $Re$FeAsO ($Re$ = rare earth) \cite{kamihara}, 122-type
$A$Fe$_2$As$_2$ ($A$=Ba, Sr, or Ca) \cite{rotter}, 111-type $B$FeAs
($B$ = alkali metal) \cite{wang}, and 11-type tetragonal
$\alpha$-FeSe(Te) \cite{hsu}. All these compounds share the same
structural feature that there exist the robust tetrahedral layers
where Fe atoms are tetragonally coordinated by pnicogen or chalcogen
atoms and the superconduction pairing may happen. A universal
finding is that all these compounds are in a collinear
antiferromagnetic (AFM) order below a tetragonal-orthorhombic
structural transition temperature \cite{cruz,dong} except for
$\alpha$-FeTe that is in a bi-collinear AFM order below a
tetragonal-triclinic structural transition temperature
\cite{ma,bao,shi}.

The above finding brings out a serious issue regarding the mechanism
behind the structural and antiferromagnetic transitions, and their
underlying relationship. There are basically two contradictive views
upon this issue. The first one \cite{mazin} is based on itinerant
electron picture, which thinks that there are no local moments and
the collinear antiferromagnetic order is induced by the Fermi
surface nesting that is also responsible for the structural
transition due to breaking the four-fold rotational symmetry. On the
contrary, the second one is based on local moment picture. The
$J_1$-$J_2$ Heisenberg model was, phenomenologically \cite{yildirim}
and from strong electron correlation limit \cite{si} respectively,
proposed to account for the issue. At the same time, we proposed
\cite{ma1} the fluctuating Fe local moments with the As-bridged
antiferromagnetic superexchange interactions as the driving force
upon the two transitions, effectively described by the $J_1$-$J_2$
Heisenberg model as well.

Our proposal embodies the twofold meanings shown by the calculations
\cite{ma1}: (1) there are localized magnetic moments around Fe ions
and embedded in itinerant electrons in real space; (2) it is those
bands far from rather than nearby the Fermi energy that determine
the magnetic behavior of pnictides, namely the hybridization of Fe
with the neighbor As atoms plays a substantial role. Here the
formation of a local moment on Fe ion is mainly due to the strong
Hund's rule coupling on Fe $3d$-orbitals \cite{ma1}. In this sense,
our proposal can be considered as the Hund's rule correlation
picture. We emphasize again that the Arsenic atoms play a
substantial role in our physical picture. Subsequently we
successfully predicted from the calculations\cite{ma}, based on this
Hund's rule correlation picture, that the ground state of
$\alpha$-FeTe is in a bi-collinear antiferromagnetic order, which
was confirmed by the later neutron scattering experiment \cite{shi}.

At ambient pressure, like other pnictides, CaFe$_2$As$_2$ undergoes
a transition from a nonmagnetically ordered tetragonal (T) phase to
a collinear AFM orthorhombic (O) phase below 170K \cite{exp-cafm}
(\textbf{T}$_{T\rightarrow O}$). Nevertheless, the neutron
diffraction measurements\cite{prb08,prb09} further found that, at
50K, CaFe$_2$As$_2$ experiences another exceptional transition from
a collinear AFM orthorhombic phase to a new nonmagnetically ordered
but collapsed tetragonal (cT) phase with a dramatic decrease in both
the unit cell volume and the $c/a$ ratio when the applied pressure
is larger than 0.35 GPa (\textbf{T}$_{O\rightarrow cT}$). The
corresponding phase diagram is schematically shown in Fig. 1(a).
Thus how to understand this phase diagram and accompanied magnetism,
especially the two nonmagnetically ordered tetragonal phases,
becomes an outstanding problem.

In this Letter, we show that the exceptional collapsed tetragonal
phase and related structure transitions, accompanied by the magnetic
transitions, found experimentally in CaFe$_2$As$_2$ can be well
understood within the proposed Hund's rule correlation picture. The
importance of the fluctuating local moments in the structural
transitions is elucidated. Moreover, we predict a pressure- and
temperature-induced phase diagram for BaFe$_2$As$_2$ in the same
framework.

The previous calculations \cite{dft-cafm} show that the ground state
of CaFe$_2$As$_2$ is in an orthorhombic structure with a collinear
AFM order at low temperature and low pressure, in an excellent
agreement with the experiments. It was naturally proposed, from the
Hund's rule correlation picture, that the high-temperature
tetragonal phase of CaFe$_2$As$_2$ is a {\emph paramagnetic} (PM)
phase with local magnetic moments around Fe atoms randomly
orientated due to thermal fluctuations, which is supported by the
universal linear-temperature dependence of static magnetic
susceptibility in iron pnictides \cite{gmzhang} and our calculation
illustrated below. Here, we further propose that the
pressure-induced exceptional collapsed tetragonal phase of
CaFe$_2$As$_2$ is a {\emph nonmagnetic} (NM) phase, in which the
Hund's rule in Fe atoms is overcome so that the Fe magnetic moments
are quenched. The corresponding calculations reported below confirm
our proposal and well describe the exceptional phase diagram of
CaFe$_2$As$_2$.

Our density functional theory (DFT) calculations were done using the
general gradient approximation (GGA) for exchange-correlation
potentials\cite{pbe}, the projector augmented wave method
\cite{paw}, and a plane wave basis set up to 600 eV as implemented
in the Vienna {\it ab-initio} simulation package\cite{vasp}. The
Gaussian broadening technique was used and a $k$-mesh of $16\times
16\times 4$ or $12\times12\times4$ was adopted to sample the
Brillouin zone of $1\times1$ or $\sqrt{2}\times\sqrt{2}$ supercell,
respectively. The shape of the lattice and the internal coordinates
of all ions were fully optimized at a given cell volume. A series of
volumes, differing in $0.5\%$ in each dimension from one to another,
were calculated for each state, as reported in Figs.
\ref{fig:afm-nm} and \ref{fig:Ca-PM2NM}.

In calculation, it is very difficult to directly simulate a
paramagnetic phase. Nevertheless, the checkerboard-AFM and
paramagnetic phases share the following important features: (1)
local moments around Fe atoms, (2) zero net magnetic moments in a
unit cell, and (3) the same magnetic symmetry. Moreover, the effect
of spin fluctuations on static properties in the paramagnetic phase
may be accounted for by the checkerboard-AFM state. Thus as long as
we study the static properties rather than the low-energy dynamics,
the checkerboard-AFM state is feasible to properly model the
paramagnetic phase. Accordingly, the orthorhombic collinear AFM
(c-AFM), tetragonal checkerboard-AFM namely Neel order (denoted as
Neel-AFM), and tetragonal NM states are considered to represent the
low-temperature orthorhombic (O) phase, the high-temperature
tetragonal (T) phase, and the pressure-induced collapsed tetragonal
(cT) phase in CaFe$_2$As$_2$, respectively.

\begin{figure}[]
\includegraphics[bb= 0 0 86 125]{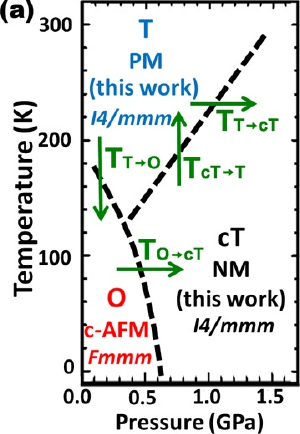}
\includegraphics[bb= 0 0 80 122]{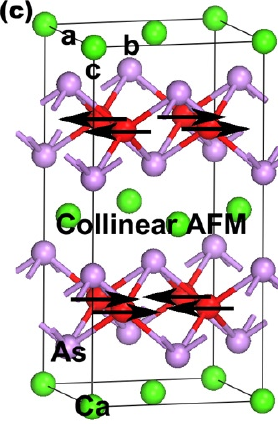}
\includegraphics[bb= 0 0 71 122]{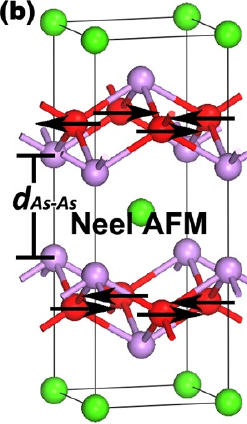}
\caption{(a) Sketch of the phase diagram of CaFe$_2$As$_2$. Four
transitions are marked: transition from tetragonal (T) phase  to
orthorhombic (O) phase when the temperature decreases below 170K at
ambient pressure ({\bf T}$_{T\rightarrow O}$), in which a long-range
collinear antiferromagnetic order is established; this magnetic
order disintegrates in collapsed tetragonal (cT) phase by applying a
pressure({\bf T}$_{O\rightarrow cT}$); phase T recovers from cT when
increasing the temperature ({\bf T}$_{cT\rightarrow T}$); and phase
T transforms again to cT by applying a large pressure ({\bf
T}$_{T\rightarrow cT}$). (b) Supercells of phases T or cT in ${I
4/mmm}$ symmetry, and (c) supercell of phase O in ${Fmmm}$ symmetry.
Green, pink, and red spheres represent Ca, As and Fe atoms,
respectively. Black arrows denote the magnetic moments of Fe. The
magnetic moments are quenched in phase cT.} \label{fig:str}
\end{figure}

{\it Phase transition from O to cT} is so unusual that phase O
collapses to phase cT under a small external pressure. Figure
\ref{fig:afm-nm} shows the evolution of the total energy with the
unit cell volume ($E-V$ curve) in the c-AFM and NM states. The
measured structural parameters for phases O and cT at 50~K were well
reproduced, within errors of 1-2$\%$, by the DFT calculations in the
c-AFM and NM states, respectively, while the Neel-AFM state cannot
reproduce. One sees that the volume decrease is accompanied by a
loss of magnetic moment in the c-AFM state, as shown by the red
numbers in Fig. \ref{fig:afm-nm}. Nevertheless, in the transition
region, the magnetic moment is still large, roughly 1.6-1.7
$\mu_{B}$, which indicates that the full suppression of magnetic
moments, after the c-AFM to NM transition, cannot be simply ascribed
to pressure-induced full delocalization of localized electrons
solely.

\begin{figure}[]
\includegraphics[bb= 0 0 198 146]{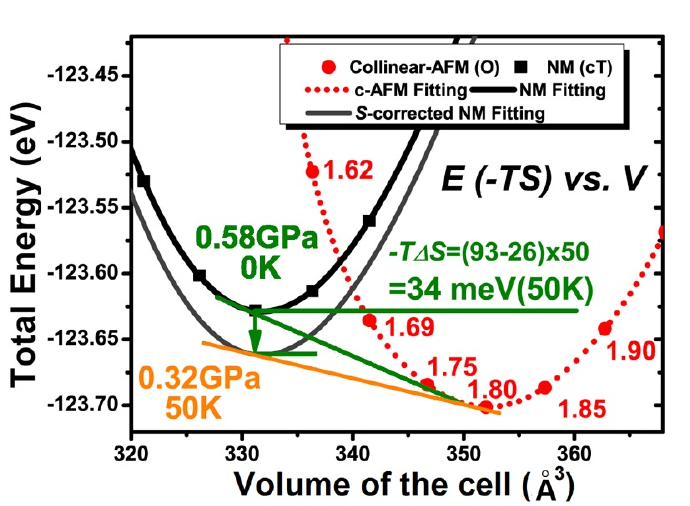}
\caption{$E-V$ curves of the NM and c-AFM states modeling phases cT
and O, respectively. Magnetic moment of each data point in phase O
is indicated by a red number. A downshift of 34 meV of the NM curve
is made according to $-T\Delta S_{cT,O}$ at 50K. This predicts
transition pressures of 0.58 GPa and 0.32 GPa at 0K and 50K,
respectively.} \label{fig:afm-nm}
\end{figure}

Gibbs free energy $G$ determines the stability of phases, which is
defined as $G=E-TS+pV$, where $E$, $T$, $S$, $p$, and $V$ stand for
internal energy, temperature, entropy, pressure, and volume,
respectively. The transition pressure of \textbf{T}$_{O\rightarrow
cT}$ can be thus worked out by $p =((E_{NM}-E_{c-AFM}) -
T(S_{NM}-S_{c-AFM}))/(V_{c-AFM}-V_{NM}$), whereas the entropy terms
are negligible at extremely low temperature, e.g. 4K. It turns out
that the transition pressure from c-AFM to NM state is given by the
slope of the tangent line between the NM and c-AFM curves in Fig.
\ref{fig:afm-nm}, i.e. 0.58 GPa at 0K, which excellently agrees with
the experimental result of 0.6 GPa at 4K\cite{prb09}. In contrast,
this value is smaller by one order of magnitude than another
calculated value of 5.25 GPa working with the Fermi surface nesting
picture\cite{zhang2010}. Meanwhile, a dramatic cell volume deduction
of about 19 \AA$^3$ is found in our calculations, consistent with
the experimental value of 14.23~\AA$^{3}$ at 50K.

At a finite temperature, the entropy terms need to be considered.
The entropy change from phase cT to O ($\Delta S_{cT,O}$) is
estimated as 0.67 meV/K\cite{entropy}. This lowers the $G$ of phase
cT by 34 meV at 50 K, as illustrated by the gray curve in Fig.
\ref{fig:afm-nm}. Following the same procedure, another tangent line
in the figure gives a transition pressure of 0.32 GPa and a similar
volume deduction, again excellently consistent with the measured
pressure in between 0.24 and 0.35 GPa at 50 K\cite{prb08}.

\begin{figure}
\includegraphics[bb= 0 0 198 146]{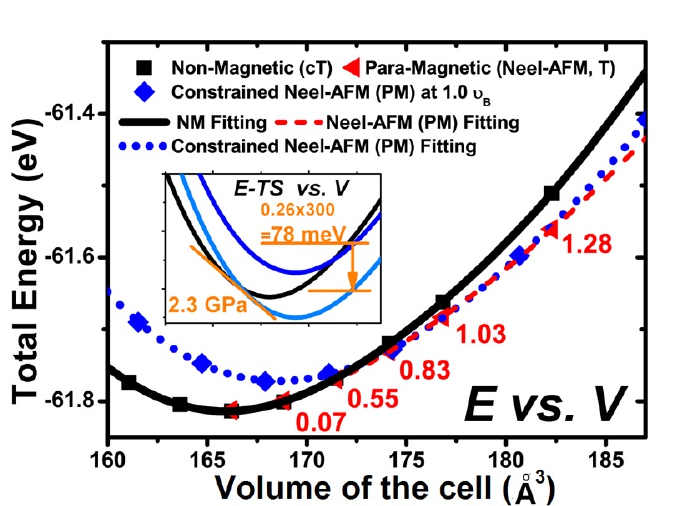}
\caption{$E-V$ curves of the NM (black solid), Neel-AFM (red
dashed), and constrained Neel-AFM (blue dotted) states. Magnetic
moment of each data point in Neel-AFM state is indicated by a red
number. A downshift of 78 meV of the constrained Neel-AFM curve,
corresponding to the entropy difference between phases T and cT at
300K, is shown in the inset, which gives a transition pressure of
2.3 GPa.} \label{fig:Ca-PM2NM}
\end{figure}

{\it Phase transitions between T and cT} -- Figure
\ref{fig:Ca-PM2NM} shows the $E-V$ curves of the Neel-AFM and NM
states. We find that the NM state has a lower energy than the
Neel-AFM state at the equilibrium. The Neel-AFM state is, however,
more stable than the NM state at larger volumes, for example, at a
volume of 176.83 \AA$^{3}$ close to the experimental value of phase
T at 0.63 GPa and 200 K\cite{prb08}. The corresponding local moment
increases to 1.03 $\mu_{B}$. And the nearest neighbors Fe-Fe
separation in the Neel-AFM state is only 0.03~\AA~ shorter than the
experimental value while 0.14~\AA~ in the NM state.

There are two phase transitions defined with phases T and cT, i.e.
transitions \textbf{T}$_{cT\rightarrow T}$ and
\textbf{T}$_{T\rightarrow cT}$ as marked in Fig. \ref{fig:str}. To
well understand these two phases, another spin-polarized calculation
with magnetic moments constrained at 1.0 $\mu_{B}$ was performed, as
reported by the blue dotted curve in Fig. \ref{fig:Ca-PM2NM}.
Because of containing an extra freedom degree of magnetic moments,
the paramagnetic tetragonal phase T represented by the Neel-AFM
state has a larger entropy $S$ than the collapsed tetragonal phase
cT represented by the NM state. This lowers Gibbs free energy in
phase T more than in phase cT when increasing temperature. This
change is presented by the inset of Fig. \ref{fig:Ca-PM2NM}, in
which the blue curve (constrained Neel-AFM) goes lower at a finite
temperature by $-T\Delta S$. As a consequence, the system takes
transition from phase cT to phase T when the entropy change
overcomes the volume expansion by increasing temperature, meanwhile
the magnetic moments recover. The entropy change is estimated as
$\sim 0.26$ meV/K between phases T and cT\cite{entropy}, by which a
transition pressure of 2.3 GPa for \textbf{T}$_{cT\rightarrow T}$ is
derived from the $S$-corrected $E-V$ curve ($E-TS$ versus $V$ in
Fig.\ref{fig:Ca-PM2NM} inset) at $T$=300~K, reasonably consistent
with the experimental value of 1.5 GPa\cite{prb09}.

{\it Collapse mechanism of CaFe$_{2}As_{2}$} -- As analyzed above,
the Fe magnetic moment is induced by the Hund's rule coupling on the
Fe $3d$-orbitals, which is about 0.6-0.8 eV/Fe. In transition
\textbf{T}$_{O\rightarrow cT}$, the Hund's rule coupling must be
overcome and compensated from Gibbs free energy gain since the
magnetic moments are quenched. Inspection of the calculations shows
that the spin degeneration pushes down the energy levels raised by
the Hund's rule coupling, and then strengthens the bonding between
the Fe $3d$ orbitals and As $4p$ orbitals, which makes substantial
energy gain to the NM state. Besides, there is an energy of
$\sim$0.1 eV/Fe (denoted as $E_{R}$) contributed from other sources
to compensate the Hund's rule coupling, as suggested by the total
energy difference between the non-collapsed NM state and the c-AFM
state of CaFe$_{2}$As$_{2}$ in the equilibrium. There are such
possible sources as: (1) entropy change effect of $-T\Delta S$; (2)
volume shrink effect of $p\Delta V$; and (3) formation of new bonds
that make further energy gain.

The entropy difference between the c-AFM and NM states is very small
and can be negligible in very low temperatures. As reported above,
the cell volume containing four Fe ions is dramatically reduced by
19\AA$^3$ after transition \textbf{T}$_{O\rightarrow cT}$ from the
c-AFM to the collapsed NM state. In order to compensate $E_R$ in
Gibbs free energy to favor the collapsed phase through the volume
shrink effect of $p\Delta V$ alone, an applied pressure of 3.4 GPa
would be required. However, it is noticed that CaFe$_{2}$As$_{2}$ is
so compact that the nearest As-As distance $d_{As-As}$ in the
equilibrium, as shown in Fig. \ref{fig:str}(b), is already very
close to the distance that can form an As-As covalent bond, i.e.
$\sim$2.9~\AA. Actually, the As-As bond indeed forms in the
collapsed phase \cite{prl09}. The formation of such a new bond
contributes an additional energy gain of $\sim$0.1 eV/Fe, as
estimated from the total energy difference between the collapsed and
non-collapsed NM states of CaFe$_2$As$_2$. Such an energy gain is
fairly comparable with $E_{R}$. It turns out that the c-AFM to the
collapsed NM transition can take place at a low pressure that pushes
the nearest As atoms close enough to form bonds, due to the small
size of Ca atoms and its resulting short inter-layer As-As distance.
The collapsed phase is thus closely related to the formation of the
As-As bonds. In most pnictides, the nearest As-As distance in
equilibrium is much larger than the one in CaFe$_2$As$_2$,
differences in phase diagrams are thus expected, as found in
BaFe$_2$As$_2$ reported below.

\begin{figure}
\includegraphics[bb= 0 0 156 114]{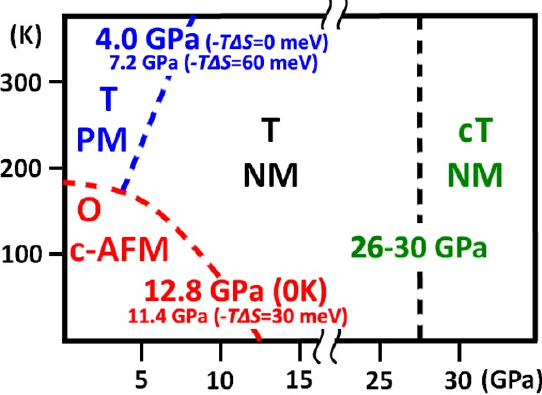}
\caption{Sketch of the predicted phase diagram of
BaFe$_{2}$As$_{2}$.} \label{fig:ba-diagram}
\end{figure}

{\it Prediction on phase diagram of BaFe$_{2}$As$_{2}$} -- Within
the Hund's rule correlation picture, we predict a phase diagram of
BaFe$_{2}$As$_{2}$, as shown in Fig. \ref{fig:ba-diagram}, which
looks similar in shape to that of CaFe$_2$As$_2$. However, the
calculation shows that the realization of an effective As-As bond
needs a huge pressure (over 20 GPa) due to a large $d_{As-As}$ in
BaFe$_2$As$_2$. Such a large pressure already provides a sufficient
Gibbs free energy gain through $p\Delta V$ to compensate the Hund's
rule coupling, resulting in a nonmagnetic state, not necessarily
with an extra energy gain from the As-As bonding in collapsed
structures. Consequently, there is no c-AFM directly to collapsed NM
transition. Instead, there are two transitions with increasing
pressure, first from the c-AFM to a normal NM state, then further to
a collapsed NM state. The corresponding transition pressures are
predicted following the same way adopted for CaFe$_2$As$_2$. They
are, as shown in Fig. \ref{fig:ba-diagram}, (1) the paramagnetic
tetragonal to the NM tetragonal phase transition at 4.0-7.2 GPa
depending on temperature $T$; (2) the c-AFM orthorhombic to the NM
tetragonal phase transition around 12.8 GPa at very low $T$; (3) the
NM tetragonal to the collapsed NM tetragonal phase transition at
26-30 GPa.

In summary, with the Hund's rule correlation picture, the
exceptional phase transitions accompanied with the magnetic
structures and properties in CaFe$_{2}$As$_{2}$ are well understood,
as demonstrated using the density functional theory calculations.
The calculated transition pressures of 0.58 (0.32) GPa at 0 (50) K
for the transition from the collinear antiferromagnetic orthorhombic
phase to the nonmagnetic collapsed tetragonal phase, and 2.3 GPa at
300 K for the transition from the paramagnetic tetragonal phase to
the nonmagnetic collapsed tetragonal phase, are highly consistent
with the experimental values 0.6 (0.25-0.35) GPa at 4 (50) K and 1.5
GPa at 300 K, respectively. Different from CaFe$_2$As$_2$, a
nonmagnetic non-collapsed tetragonal phase is predicted to appear in
the phase diagram of BaFe$_2$As$_2$ due to the larger atomic size of
Ba than that of Ca, separating the nonmagnetic collapsed tetragonal
phase from the collinear antiferromagnetic orthorhombic phase and
paramagnetic tetragonal phase. The corresponding transition
pressures are predicted as well and reported in phase diagram Fig.
\ref{fig:ba-diagram}.


This work was supported by National Natural Science Foundation of
China and by National Program for Basic Research of MOST, China.


\begin{thebibliography}{99}
\bibitem{entropy}
The transition between phases PM T and NM cT occurs at 0.63 GPa,
200~K\cite{prb08}. According to these values and the calcculated
$E-V$ curves, $\Delta S_{T,cT}(S_{Neel-AFM}-S_{NM})$ is derived as
0.26~meV/K. Likewise, $\Delta S_{T,O}$=0.93 meV/K between phases T
and O is derived. These lead $\Delta S_{cT,O}$=0.67 meV/K between
phases cT and O, assuming $\Delta S$ unchanged in the considered
tmeperature region.
\end{thebibliography}
\end{document}